\documentclass{PoS}

\newcommand{\aj}{{\it AJ}}
\newcommand{\apj}{{\it ApJ}}
\newcommand{\apjl}{{\it ApJ}}
\newcommand{\aap}{{\it A\&A}}
\newcommand{\mnras}{{\it MNRAS}}
\newcommand{\nat}{{\it Nature}}
\newcommand{\actaa}{{\it Acta Astron.}}

\newcommand{\swift}{{\it Swift}}
\newcommand{\msun}{\mbox{M$_\odot$}}
\newcommand{\msunyr}{\mbox{\msun\,yr$^{-1}$}}
\newcommand{\arcsec}{{$^{\prime\prime}$}}
\newcommand{\lt}{{$<$}}

\usepackage{graphicx,natbib,amsmath,amssymb}

\bibpunct[, ]{[}{]}{,}{n}{}{,}
\renewcommand{\citet}{\citep}

\title{TOUGH: Observational aspects of gamma-ray burst host galaxies}

\ShortTitle{TOUGH}

\author{\speaker{Jens Hjorth},$^a$ 
Daniele Malesani,$^a$
Andreas O. Jaunsen,$^b$
Andrew J. Levan,$^c$
\mbox{Bo Milvang-Jensen,$^a$}
Darach Watson,$^a$
Javier Gorosabel,$^d$
Johan P. U. Fynbo,$^a$
Micha{\l}~J.~Micha{\l}owski,$^e$
Nial R. Tanvir,$^f$
P\'all Jakobsson,$^g$
Palle M\o ller,$^h$
\mbox{Steve Schulze$^g$} and
Thomas Kr\"uhler$^a$\\
\llap{$^a$}
        Dark Cosmology Centre, Niels Bohr Institute, University of Copenhagen\\
	DK-2100 Copenhagen, Denmark\\
\llap{$^b$}
	Institute of Theoretical Astrophysics, University of Oslo\\
	PO Box 1029 Blindern, N-0315 Oslo, Norway\\
\llap{$^c$}
	Department of Physics, University of Warwick\\
	Coventry CV4 7AL, UK\\
\llap{$^d$}
        Instituto de Astrof\'\i sica de Andaluc\'\i a (IAA-CSIC)\\
	P.O. Box 03004, E-18080 Granada, Spain\\
\llap{$^e$}
        SUPA, Institute for Astronomy, University of Edinburgh, 
	Royal Observatory\\
        Edinburgh, EH9 3HJ, UK\\
\llap{$^f$}
        Department of Physics and Astronomy, University of Leicester\\
	University Road, Leicester LE1 7RH, UK\\
\llap{$^g$}
	Centre for Astrophysics and Cosmology, Science Institute,
	University of Iceland\\
	Dunhagi 3, 107 Reykjav\'ik, Iceland\\
\llap{$^h$}
        European Southern Observatory\\
	Karl-Schwarzschild-Str.\ 2, D-85748 Garching by M\"unchen, Germany\\
E-mail: \email{jens@dark-cosmology.dk},
        \email{malesani@dark-cosmology.dk},
        \email{ajaunsen@gmail.com},
        \email{A.J.Levan@warwick.ac.uk},
        \email{milvang@dark-cosmology.dk},
        \email{darach@dark-cosmology.dk},
        \email{jgu@iaa.es},
        \email{jfynbo@dark-cosmology.dk},
        \email{mm@roe.ac.uk},
        \email{nrt3@leicester.ac.uk},
        \email{pja@hi.is},
        \email{pmoller@eso.org},
        \email{sts30@hi.is},
        \email{tom@dark-cosmology.dk}
	}

\abstract{GRB-selected galaxies are broadly known to be faint, blue, young,
star-forming dwarf galaxies. This insight, however, is based in part on
heterogeneous samples of optically selected, lower-redshift galaxies.
To study the statistical properties of GRB-selected galaxies we here introduce 
The Optically Unbiased GRB Host (TOUGH) complete sample of 69 X-ray selected 
\swift\ GRB host galaxies spanning the redshift range 0.03--6.30 and summarise 
the first results of a large observational survey of these galaxies.}

\FullConference{Gamma-Ray Bursts 2012 Conference -GRB2012,\\
		May 07-11, 2012\\
		Munich, Germany}

\begin{document}

\section{Introduction}

Gamma-ray bursts (GRBs) arise from the deaths of short-lived stars
\citep{2003Natur.423..847H}. Hence, their host galaxies bear important 
information about the nature of GRB progenitors and furthermore act as 
tracers of star formation at a very wide range of redshifts. 

The host galaxy of the nearest known GRB host galaxy, that of 
GRB 980425/SN 1998bw at $z=0.0085$ \citep{1998Natur.395..670G}, may be seen
as representative of this population: it is a dwarf (SMC sized) star-forming 
galaxy harbouring a highly star-forming region 
\citep{2000ApJ...542L..89F,2008ApJ...672..817M}.
Indeed, in the {\it BeppoSAX/HETE-II} era, it was established that GRB host 
galaxies are typically sub-luminous, blue, young, high specific 
star-formation rate systems
\citep{1999ApJ...519L..13F,2003A&A...400..499L,2004A&A...425..913C}, with
the GRBs originating from the UV/blue light of the galaxies
\citep{2002AJ....123.1111B,2006Natur.441..463F}.
The metallicities were found to be preferentially low
\citep{2006AcA....56..333S,2008AJ....135.1136M}, providing an interesting
constraint on progenitor models of GRBs, consistent, according to the
mass-metallicity relation, with the finding that the stellar masses
are generally low \citep{2010ApJ...721.1919C,2009ApJ...691..182S}.
Early sub-mm (SCUBA) observations revealed a rare population of low-redshift 
($z\sim 1$) blue galaxies \citep{2003ApJ...588...99B,2004MNRAS.352.1073T}, 
with somewhat elevated dust temperatures \citep{2009ApJ...693..347M}.

The above results are based on heterogeneous samples of mostly lower-redshift
galaxies, e.g., relying on the existence of a bright optical afterglow 
for localisation and redshift determination.

At higher redshifts, GRB host galaxies are very faint, consistent with their 
being dwarf galaxies. For example, at $z=3.2$ (the redshift being determined 
via afterglow absorption spectroscopy) GRB 020124 
\citep{2002ApJ...581..981B,2003ApJ...597..699H} and GRB 060526 
\citep{2010A&A...523A..70T} were found to be hosted by galaxies with 
$R>29.5$ and $R>28.5$, respectively. Remarkably, there are still no 
high-redshift ($z>5$) GRB host galaxies robustly detected in emission 
\citep{2012arXiv1201.6074T},
consistent with their being sub-$L^*$ galaxies.

However, more recent results have demonstrated that the host galaxies of 
dark GRBs are more chemically evolved and have higher masses 
\citep{2011A&A...534A.108K}, complicating the early simple picture of GRB 
host galaxies being sub-luminous, low-metallicity, highly star-forming 
blue galaxies (see also the contributions by Kr\"uhler, Levesque, and
Perley to these proceedings).

\section{TOUGH}

\begin{figure}
\includegraphics[width=.9\textwidth]{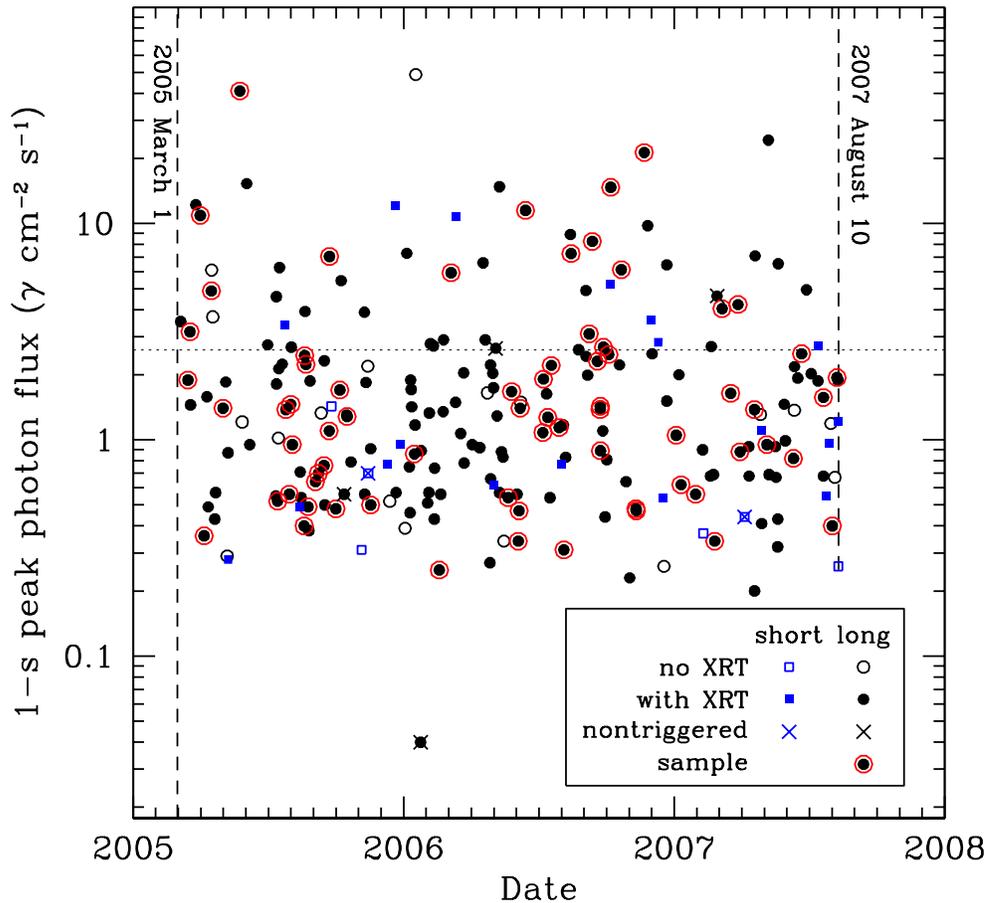}
\caption{1-s \swift/BAT peak flux (15--150 keV) for all \swift{} GRBs observed
during the time span of the TOUGH survey (bracketed by vertical dashed lines). 
Empty circles: long  bursts without an X-ray afterglow. Filled circles: long 
GRBs with \swift/XRT detection. Filled, encircled circles: GRBs obeying all 
TOUGH sample selection  criteria. Squares: GRBs classified as short 
($T_{90}<2$~s). Crosses: nontriggered GRBs. There seems to be a constant and 
uniform detection level as a function of time. The horizontal line is the 
limiting peak flux limit adopted by \citet{2012ApJ...749...68S}. 
From \citet{2012arXiv1205.3162H}.
}
\label{fig1}
\end{figure}

Taking advantage of the launch of the \swift\ mission 
\citep{2004ApJ...611.1005G}, The Optically Unbiased GRB Host (TOUGH) survey 
\citep{2012arXiv1205.3162H} was designed to define a large, carefully 
selected, homogeneous sample of 69 gamma-ray and X-ray selected galaxies, 
covering a wide redshift range. The selection criteria are detailed in 
\citet{2012arXiv1205.3162H}. Figure~\ref{fig1} shows the distribution of 
the peak gamma-ray flux of the TOUGH sample, suggesting a limit of about 
0.3--0.4 $\gamma$ cm$^{-2}$ s$^{-1}$, almost an order of magnitude deeper than 
the sample of \citet{2012ApJ...749...68S}.

\begin{figure}[h]
\includegraphics[width=.9\textwidth]{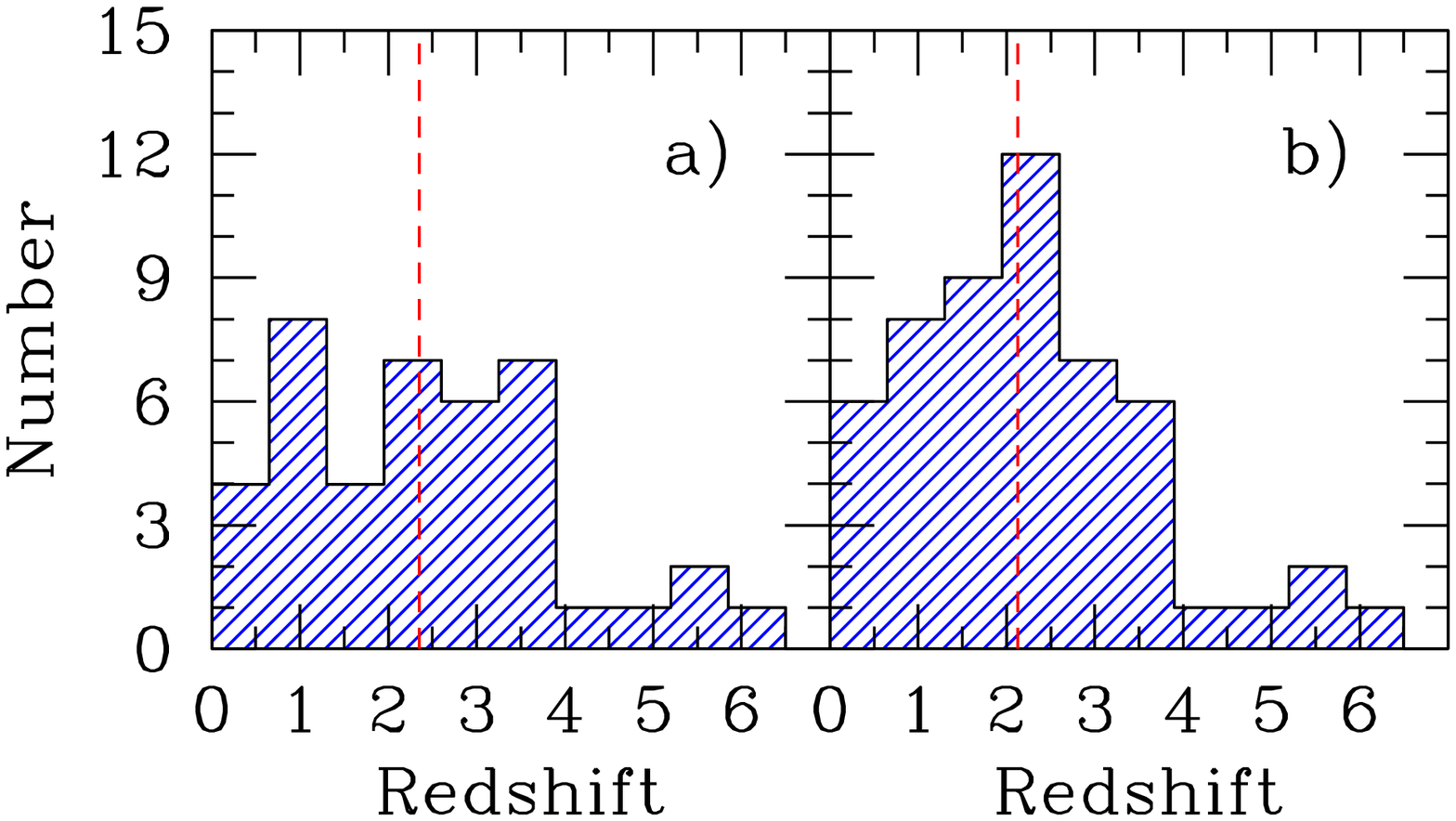}
\caption{The redshift distribution of the TOUGH sample. In both panels the
dashed vertical line indicates the median redshift. {\bf a}) Before our
survey, 38 redshifts were considered secure with a median redshift of 
$z =  2.35$. {\bf b}) Our TOUGH spectroscopic observations added 15 new 
redshifts and demonstrated that three redshifts reported in the literature 
are erroneous \citep{2012ApJ...752...62J}. Here the median redshift is 
$z = 2.14$. From \citet{2012arXiv1205.3162H}.
}
\label{fig2}
\end{figure}

Our VLT spectroscopic follow-up campaign has led to 15 new redshifts
\citep{2012ApJ...752...62J,2012arXiv1205.4036K}. The redshift distributions 
before and after our observations are shown in Figure~\ref{fig2}. 
The redshift distribution is discussed in more detail in 
\citet{2012ApJ...752...62J} and in the contribution by Jakobsson et al.\
to these proceedings. As part of our spectroscopic observations we also 
specifically targeted 20 galaxies in the redshift range 1.8--4.5 to study 
their Ly$\alpha$ emission properties \citep{2012arXiv1205.3779M}. Seven of 
these have detections of Ly$\alpha$ emission and we can therefore exclude an 
early indication that Ly$\alpha$ emission is ubiquitous among GRB hosts. 
However, we confirm that Ly$\alpha$ is stronger in GRB-selected galaxies than 
in flux-limited samples of Lyman break galaxies.

\begin{figure}[h]
\includegraphics[width=.9\textwidth]{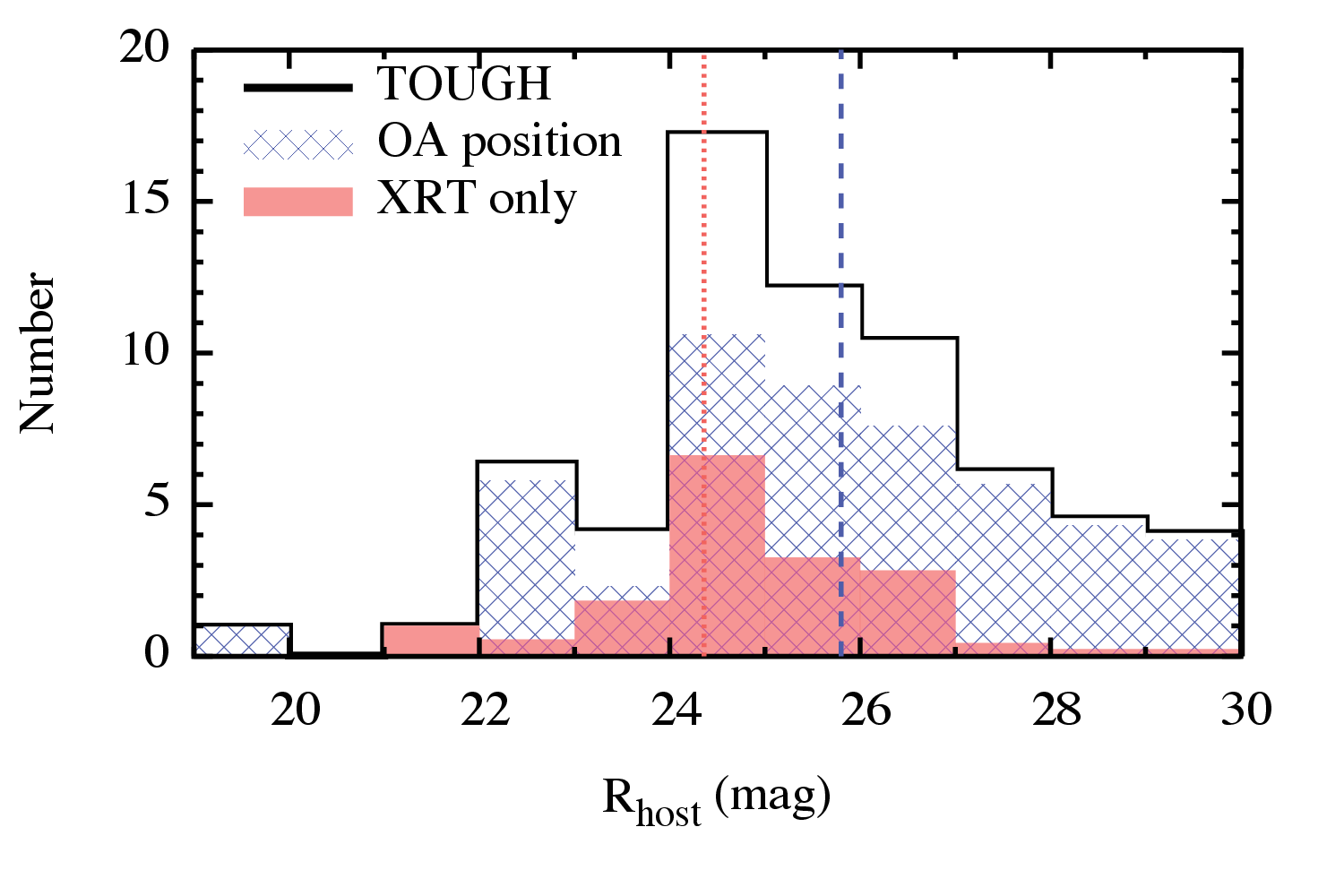}
\caption{Distributions of host galaxy $R$-band magnitudes. Subsets of galaxies 
selected with a localisation from an optical afterglow and a localisation 
from \swift/XRT only are shown as hatched and solid histograms respectively. 
The corresponding median values are shown as dashed or dotted lines. 
Adapted from \citet{2012arXiv1205.3162H}.
}
\label{fig3}
\end{figure}

The survey also consists of deep $R$- and $K$-band imaging. A key aspect
of TOUGH is that we have imaged galaxies with no optical afterglows (i.e., 
localised in the X-rays by XRT to 2\arcsec\ precision) as well as those with 
optical afterglows. The resulting distribution of $R$-band magnitudes is 
shown in Figure~\ref{fig3}. Remarkably, the host galaxies of GRBs with no 
optical afterglows are much brighter on average than those with an 
optical afterglow \citep{2012arXiv1205.3162H}.

\begin{figure}[h]
\includegraphics[width=.9\textwidth]{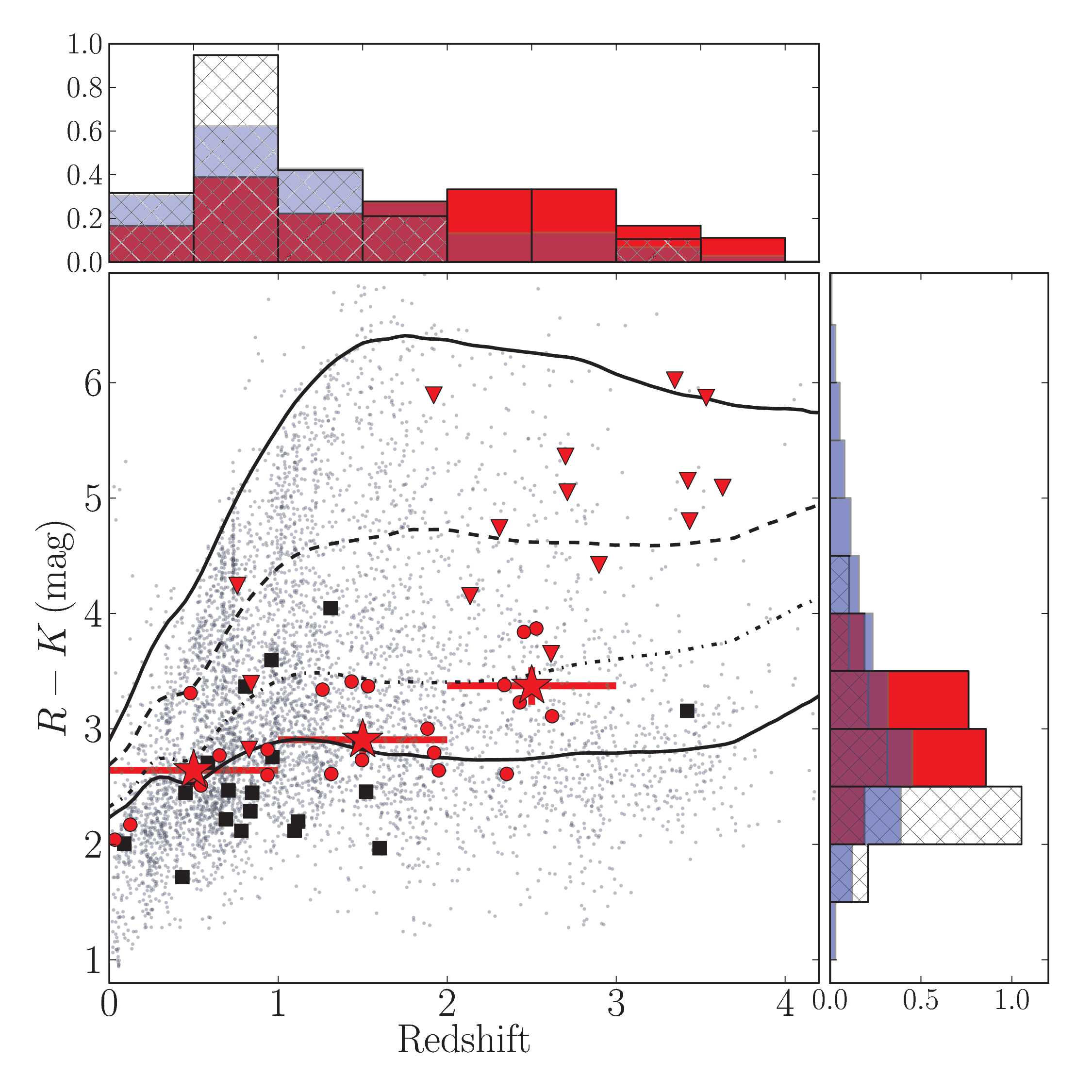}
\caption{$R-K$ colour of TOUGH GRB hosts plotted against redshift.
Red points represent TOUGH hosts detected in $R$ and $K$, while downward
triangles represent those with $K$-band upper limits. Red stars illustrate
the median $R-K$ colour over the indicated redshift interval. The data in
black squares are the GRB-host sample of \citet{2009ApJ...691..182S} and the
grey background sample are field galaxies from the FIREWORKS catalog 
\citep{2008ApJ...682..985W}. We also show redshift-dependent colours of 
different Hubble-type galaxies. These are (from red to blue $R-K$ colours): 
An old (3~Gyr) elliptical, S0 and Sbc spirals (1~Gyr) and a young (300~Myr) 
irregular galaxy at solar metallicity. These tracks have been derived from the 
PEGASE2 library integrated over the $R$ and $K$ bands. 
}
\label{fig4}
\end{figure}

The colour distribution is shown in Figure~\ref{fig4}. The median colours of 
XRT-only GRB host galaxies detected in the $K$-band are significantly redder, 
$R-K=3.7\pm0.3$, than those for which an optical afterglow was detected, 
$R-K=2.8\pm0.1$. GRBs without an optical afterglow may suffer stronger 
attenuation due to dust which may cause the optical afterglows to escape 
detection. 

\begin{figure}[h]
\includegraphics[width=.9\textwidth]{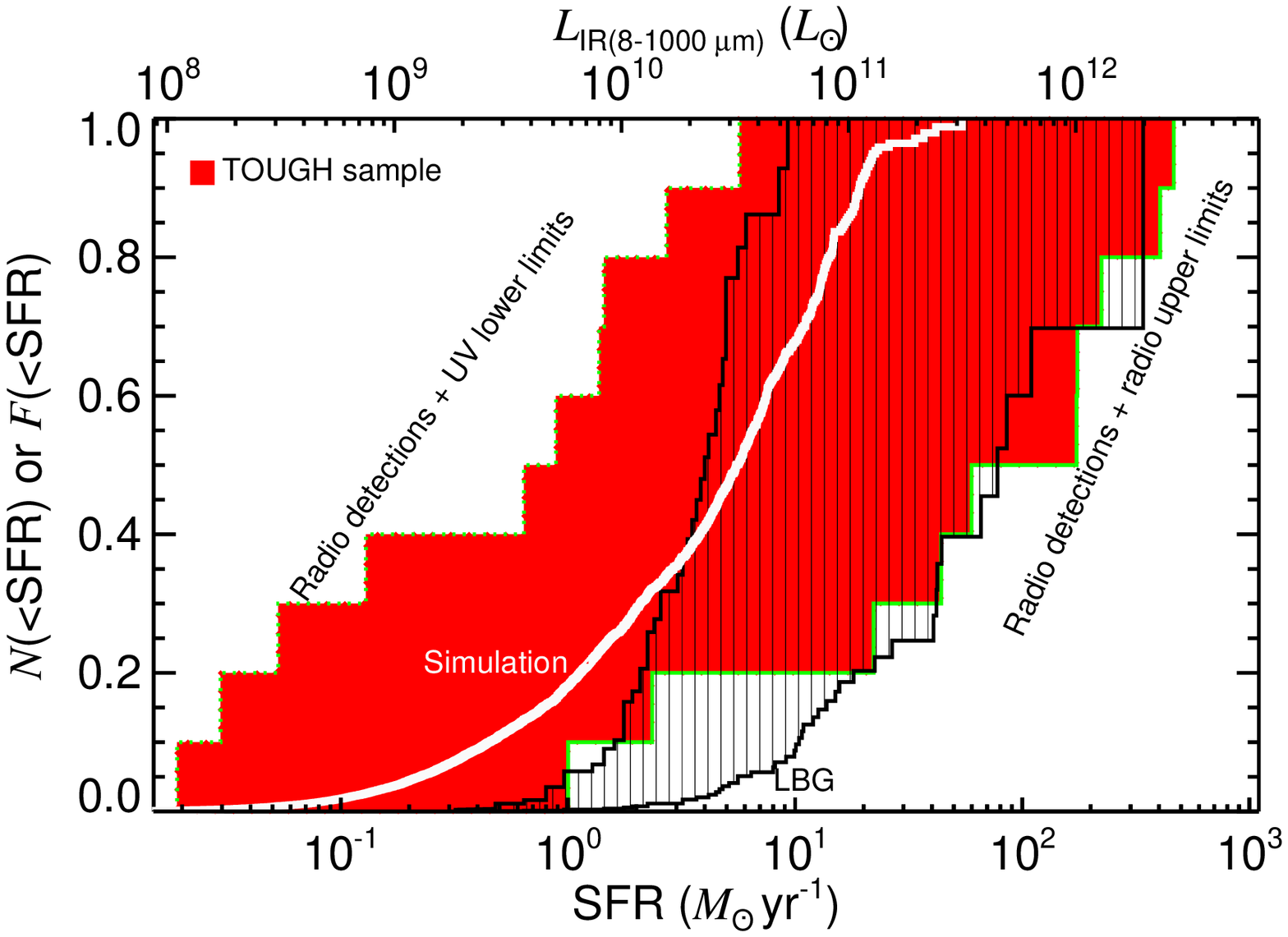}
\caption{Cumulative distribution of SFRs of GRB hosts in the $z<1$ TOUGH 
sample (red area). The high-SFR boundaries (green solid lines) are 
constructed using the detections and limits of SFR$_{\rm radio}$, whereas the 
low-SFR boundaries (green dotted lines) are constructed using the 
SFR$_{\rm UV}$ for galaxies not detected in the radio. At least $63$\% 
($>$15/24) of GRB hosts at $z\lesssim1$ have SFR$\mbox{}<100\, \msunyr$ and 
only $<8$\% ($<$2/24) could have SFR$\mbox{}>500\,\msunyr$. For comparison, 
the SFR distributions of $z=0.51$ simulated galaxies (white curve) and 
$z\sim1$ Lyman break galaxies (black curves; the right lines represent 
dust-corrected SFRs).  These distributions were weighted by SFR (so they 
reflect the fraction of total star formation in the sample contributed by 
galaxies with SFRs lower than a given SFR) to allow a comparison with the 
GRB host population, which is likely selected based on SFRs. The current SFR 
limits imply that the GRB host population is consistent with star-forming 
galaxies at similar redshifts. Adapted from \citet{2012arXiv1205.4239M}.
}
\label{fig5}
\end{figure}

In order to assess the star-formation rates (SFRs) of GRB hosts, we have
also performed radio observations of TOUGH GRB hosts at $z<1$ 
\citep{2012arXiv1205.4239M}. We did not detect any TOUGH GRB hosts, which 
indicates that their average SFR is below $\sim15$ M$_\odot~\mbox{yr}^{-1}$. 
We also found that at least $65$\% of GRB hosts at $z<1$ have 
$\mbox{SFR}<100$ M$_\odot\mbox{ yr}^{-1}$ and that at least $92$\% of them 
have $A_V<3$ mag. The distribution of SFRs and dust attenuation of GRB hosts 
at $z<1$ are consistent with those of other star-forming galaxies at similar
redshifts (Figure~\ref{fig5}). 

Papers in preparation will address the luminosity function of GRB host 
galaxies, the X-ray absorption properties, metallicities of GRB host galaxies, 
and the correlation between prompt, afterglow, and host properties. Current 
and future work will concentrate on increasing the redshift completeness of 
the sample, extending the spectral coverage to longer wavelengths (radio, 
sub-mm, with {\it Spitzer}, {\it Herschel}, ALMA and EVLA) and imaging the 
fainter galaxies with {\it HST}.  Enlarging the sample is also being 
considered. In parallel, detailed studies of TOUGH subsamples are in progress.

The TOUGH website is at: {\tt http://www.dark-cosmology.dk/TOUGH} where 
catalogs and reduced data are made publicly available.

\acknowledgments
We thank the referee, Sandra Savaglio, for helpful comments.
The Dark Cosmology Centre is funded by the Danish National Research Foundation.


\end{document}